\documentclass{aa}
\usepackage{graphicx}
\usepackage{txfonts}
\begin{document}
   \title{Coronene and pyrene (5, 7)-member ring defects}

   \subtitle{Infrared spectra, energetics and alternative formation pathways}

   \author{S. {\"O}ttl\inst{\ref{inst1}}\and S. E. Huber\inst{\ref{inst2}}\and S. Kimeswenger\inst{\ref{inst3},\ref{inst1}}\and M. Probst\inst{\ref{inst2}}
             }

   \institute{Institute for Astro- and Particle Physics, Leopold Franzens Universit{\"{a}}t Innsbruck,
              Technikerstrasse 25, 6020 Innsbruck, Austria\newline
              \email{Silvia.Oettl@uibk.ac.at}\label{inst1}
\and
Institute of Ion Physics and Applied Physics, Leopold Franzens Universit{\"{a}}t Innsbruck, Technikerstrasse 25, 6020 Innsbruck, Austria\newline
\email{S.Huber@uibk.ac.at, Michael.Probst@uibk.ac.at}\label{inst2}
\and
Instituto de Astronom{\'{i}}a, Universidad Cat{\'{o}}lica del Norte, Avenida Angamos 0610, Antofagasta, Chile\newline
\email{Stefan.Kimeswenger@gmail.com}\label{inst3}
    }

   \date{Received 16 June 2014; accepted 15 July 2014}


\abstract{Polycyclic aromatic hydrocarbons (PAHs) are nowadays known to be one of the carriers of the ubiquitous aromatic infrared (IR) bands. The IR spectra of many astrophysical objects show IR emission features derived from PAH molecules of different size. Space-based observations have shown that those IR emission features are omnipresent and can be found in most objects. However, still some of the characteristics of the emitting population remain unclear. The emission bands show details which cannot be explained so far. These unidentified IR features require further laboratory and observational investigations.
}
{We present a theoretical study of the IR spectra of PAHs containing (5, 7)-member ring defects, focusing on pyrene ($C_{16}H_{10}$) and coronene ($C_{24}H_{12}$).}
{Using density functional theory, we investigate the effects of such defects on the IR spectra of pyrene and coronene and their cations and anions. In addition, we explore parts of the potential energy surface of the neutral species and discuss alternative formation pathways.
 }
{The addition of (5, 7)-membered ring defects in pyrene and coronene results in a change of the IR spectra, both molecules loose their typical spectroscopic signature. We find shifts in the positions of the band as well as different intensities and a rise in the number of features. The boundaries in terms of the size of the PAHs exhibiting a (5, 7)-membered ring defect are studied and shown. Investigation of the minimal energy pathway leads to a result of 8.21 eV for pyrene and 8.41 eV for coronene as minimum activation barriers for the transformation from ground state to defected state. Whereas pyrene retains some of its symmetry due to the symmetry exhibited by the Stone-Wales defect itself, coronene loses much more of its symmetry. The formation of these (5,7)-ring defects in PAHs may be well supported in asymptotic giant branch stars or planetary nebulae. Those environments strongly enable the transition from the ground state to the defect state. Therefore the knowledge of the IR spectra of these molecules will support future investigations aiming for a thorough understanding of the unidentified IR emission bands.
}
   {}

   \maketitle
%

\section{Introduction}
Polycyclic aromatic hydrocarbons (PAHs) are one of the most interesting components of the interstellar medium (ISM). The infrared (IR) spectra of different astronomical objects like HII regions, photodissociation regions (PDRs) or asymptotic giant branch (AGB) stars and planetary nebulae (PNe) are dominated by emission features in the region between 3.3 - 18${\mu}$m (Peeters \cite{peeters}, Draine \& Li \cite{draine}). It is thought that PAHs are also carriers of these diffuse interstellar bands, the so called DIBs (Tielens \cite{tielens}). But different from laboratory - the isolation of a single species is not possible. The spectra contain a superposition of many different molecules. Therefore a sophisticated understanding of all possible subtypes is required.

With the Infrared Space Observatory (ISO) and the Spitzer Space Telescope, a great variety of PAH bands was discovered (e.g., Spitzer Special Edition \cite{spitzer}, Peeters et al. \cite{peeters04}, Bauschlicher et al. \cite{bausch09}). The IR spectra are dominated by prominent emission features at 3.3, 6.2, 7.7, 8.6 and 11.2 $\mu$m and many more. Those bands are known as the unidentified infrared (UIR) bands (Tielens \cite{tielens}), but the characteristics of emitting PAHs remain unclear so far. There is a clear variability in the PAH spectra from source to source and spatially within sources (Peeters \cite{peeters}). The high sensitivity of the space-based observations have shown that the IR emission features are omnipresent and can be found in most astronomical objects (Peeters \cite{peeters}).

A spectroscopic database of PAH molecules has been published by Bauschlicher et al. (\cite{bausch10}), which has recently been updated (Bauschlicher et al. \cite{bausch14}. The web site (http://www.astrochem.org/pahdb/) contains over 800 spectra of PAHs in their neutral and charged states, experimentally measured and theoretically computed. The database contains positive and negative charged molecules, PAHs with different elements substituted or metallic-cation PAH complexes, and all of the subgroups (e.g., Bakes et al. \cite{bakes}, Knorke et al. \cite{knorke}, Galue et al. \cite{galue}, Hudgins et al. \cite{hudgins}, Bauschlicher et al. \cite{bausch08}, Simon \& Joblin \cite{simon}).

The influence of size and symmetry on the IR spectra of the PAHs as well as PAHs with different elements added has been studied in the past (e.g., Pauzat \& Ellinger \cite{pauzat}, Hudgins et al. \cite{hudgins}, Galue et al. \cite{galue}, Bauschlicher et al. \cite{bausch08}, \cite{bausch09}, Ricca \& Bauschlicher \cite{ricca3}, Knorke et al. \cite{knorke}). These studies could provide many insights and a lot of knowledge about the observed PAHs was gained. The astrophysical environments were studied in detail, but still there are many unresolved questions.

PAHs are thought to be formed in the late stages of the asymptotic giant branch (AGB) phase (Allamandola et al. \cite{alla}). Especially the carbon-rich AGB stars are an ideal environment for PAH formation. During the growth of large, aromatic molecules, the (5, 7)-member ring defects play an important role.

\begin{figure}[ht!]
   \centering
   \includegraphics[width=8.8cm]{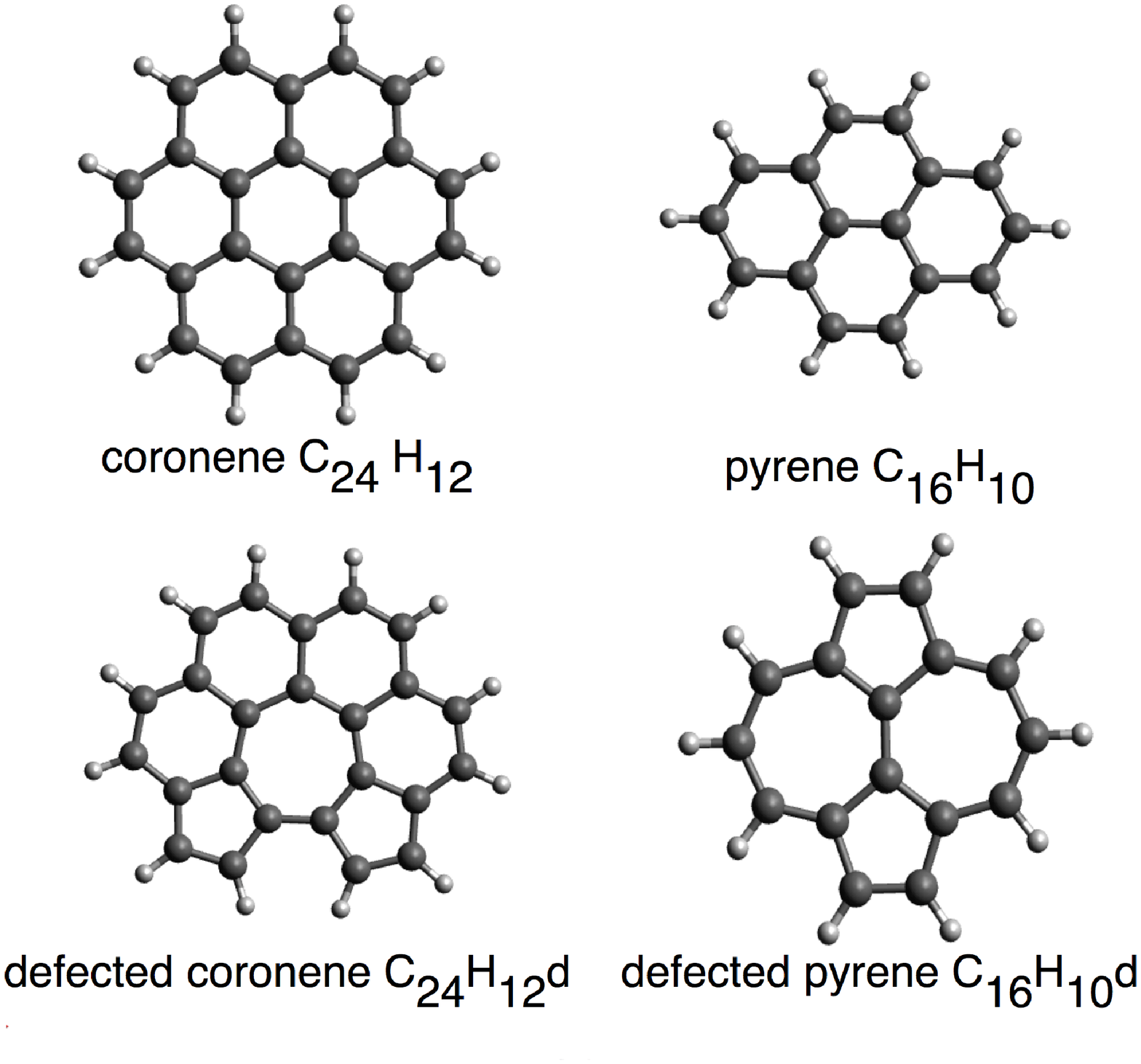}
      \caption{The PAH molecules studied in this work. Above you can see coronene and pyrene in its normal state, below there is the defected form. In $C_{24}H_{12}$d and $C_{16}H_{10}$d the 5-member rings and 7-member rings are shown.}
      \label{molecules.fig}
\end{figure}

The molecules chosen in this work are pyrene ($C_{16}H_{10}$) and coronene ($C_{24}H_{12}$), shown in Fig.\ref{molecules.fig}. Both of them are very unique for different reasons and they can be seen as limiting or prototypical cases. Coronene and pyrene are among the smallest hydrocarbons that can exhibit a (5,7)-member ring defect, pyrene is the smallest one. Previous work on (5,7)-member ring defects (Yu \& Nyman \cite{yu}) has raised the suspicion that effects of such defects might be more intriguing with decreasing size of the PAHs. To study these small systems is thought to be an important step towards a substantial understanding of the role of (5,7)-ring defects for the IR spectra of PAHs. Another peculiarity of these two molecules appears by investigating the minimal energy pathway between the ground and (5,7)-ring defected states, see section 3. Doing so, it turns out that coronene proceeds via a Stone-Wales transition (Harris \cite{harris}) from its ground to its defected state, but shows no Stone-Wales defect in the defected form. On the other hand, pyrene shows no Stone-Wales transition, but instead the defected pyrene exhibits a Stone-Wales defect. The latter is the underlying reason for the third respect in which the two studied molecules can be regarded as limiting and prototypical. Whereas pyrene retains some of its symmetry due to the symmetry exhibited by the Stone-Wales defect itself, coronene loses much more of its symmetry (see Fig.\ref{molecules.fig}).

This work is structured as follows. In section 2, we give a short description of the methodology and the computational methods used in this work. In section 3.1, we discuss the effect of (5,7)-membered ring defects on the IR absorption spectra of pyrene, coronene and their (singly charged) cations and anions. This is followed by an exploration of parts of the potential energy surface of the neutral species to investigate their thermodynamic stability in section 3.2. In section 4 we present an alternative pathway that could lead to the formation of these defected PAHs in the interstellar matter and discuss their stability, observability and detectability from an astrophysical point of view.

\section{Models and computational methods}
Pyrene ($C_{16}H_{10}$) consists in its original condition of four 6-membered rings. In the defected state, all four rings are changed into two 5-membered and two 7-membered rings. The size of the molecule stays the same. The defected state is referred as $C_{16}H_{10}$d. The normal  coronene ($C_{24}H_{12}$) consists of a 6-membered ring in the center, surrounded by a ring of six more 6-membered rings. The defect changes the ring in the center into a 7-membered ring and two adjacent rings are changed into 5-membered rings. Coronene with defect is denoted as $C_{24}H_{12}$d.

The geometry optimizations and harmonic frequency calculations were performed using the B3LYP density functional, which uses Becke's three parameter hybrid exchange functional (Becke \cite{becke}) and a correlation functional of Lee, Yang and Parr comprising both local and non-local terms (Lee et al. \cite{lee}), in conjunction with the 4-31G basis set (Frisch et al. \cite{frisch}). It has been shown previously (Langhoff \cite{langhoff}; Bauschlicher \& Langhoff \cite{bausch}; Boersma et al. \cite{boersma}) that this combination of functional, basis set and a single scale factor of 0.958 leads to harmonic frequencies that are in excellent agreement with experiments. IR spectra, calculated using this level of theory, of molecules for which no experimental data exists are at least consistent with the data of other PAH molecules of the same size and charge (Ricca et al. \cite{ricca}). In the cases of the (5,7)-ring defected molecules no experimental data exists for comparison. Furthermore, it has been shown in previous work that the inclusion of diffuse functions is not required even for the investigation of anions (Ricca et al. \cite{ricca}). However, considering intensities, it has been reported that B3LYP might yield erroneous results if there are two electronic states with essentially the same energy (Ricca et al. \cite{ricca}). In such cases it is known that HF calculations might favor an asymmetric wavefunction, which can lead to the consequence that hybrid functionals are more likely to have problems than generalized gradient approximation (GGA) functionals due to the inclusion of HF exchange (Sherrill \cite{sherrill}). For this reason we repeated our harmonic frequency calculations with the GGA functional of Becke and Perdew, i.e. BP86 (Becke \cite{becke88}; Perdew \cite{perdew}) using a single scale factor of 0.986. However, when B3LYP encounters no problems, it is usually more accurate than GGA functionals. Hence, we report results using the BP86 functional only in those cases when the B3LYP results deviate strongly in terms of intensities. Otherwise, the B3LYP data are taken.

All spectra are IR absorption spectra. Actually, the observations of the PAHs in space deal with the emission lines of the molecules. In order to be able to compare our theoretical spectra with the observed spectra, different line widths are taken for the different bands, in a similar manner as in Ricca et al. \cite{ricca}. The synthetic IR spectra are generated via folding the frequencies and intensities with a normalized Gaussian lineshape function. Usually a FWHM of 30 cm$^{-1}$ is taken as the natural line width of emission lines of molecules in the ISM (Bauschlicher et al. \cite{bausch09}). For the bands shorter than 10 $\mu$m, we have used this value (see Fig. \ref{69.fig}). To get better resolved spectra, we adopted a value of 10 cm$^{-1}$ for the line with in the region between 10 and 15 $\mu$m (see Fig. \ref{1015.fig}). It is thought that this value is more consistent with current observations as well as theoretical calculations (Bauschlicher et al. \cite{bausch09}). To continue to match the observations (Van Kerckhoven et al. \cite{kerck}) in the 15 - 20 $\mu$m bands, a FWHM of 5 cm$^{-1}$ is used there (see Fig. \ref{1520.fig}). For longer wavelengths, the FWHM will decrease even more. We adopt a value of 1 cm$^{-1}$ for all wavelengths above 20 $\mu$m (see Fig. \ref{20100.fig}). This is sufficient for our purposes in this region, as we want to show only the impact of the defects in the spectra.

The harmonic frequency calculations provide as unit for the synthetic spectra cm$^{-1}$ on the x-axis. Since the astrophysical observations are usually presented in $\mu$m, we converted the x-axis to $\mu$m in order to compare the results. The y-axis indicates the integrated intensities, given as a cross section in units of 10${^5}$ cm${^2}$ mol${^{-1}}$.

In order to explore the parts of the potential energy surface connecting the ground state geometries with those corresponding to the (5,7)-ring defected states we used the STQN method, which uses a quadratic synchronous transit approach to find the transition state and then completes the optimization using a quasi-Newton algorithm (Peng \& Schlegel \cite{peng}; Peng et al. \cite{peng96}). Subsequently, we performed frequency calculations to verify that we found a transition state of first order (yielding one and only one imaginary frequency) and then the reaction path was followed by integrating the intrinsic reaction coordinate (IRC) (Fukui \cite{fukui}) to verify that the transition state actually connects the ground and defected states. To assess the energetics more accurately we performed these calculations using again the B3LYP functional but now in conjunction with Dunning's cc-pVTZ basis set (Dunning \cite{dunning}). The energies of the ground, (5,7)-ring defected and transition states have then been calculated even more accurately using the G4(MP2) extrapolation approach (Curtiss et al. \cite{curtiss}) including a zero-point energy correction. These type of calculations have, however, only been performed for the neutral species.

All electronic structure calculations have been performed with the Gaussian 09 software (Frisch et al. \cite{frisch09}).

\section{Results}
\subsection{IR spectra}
\begin{figure*}[!ht]
\centering
\includegraphics{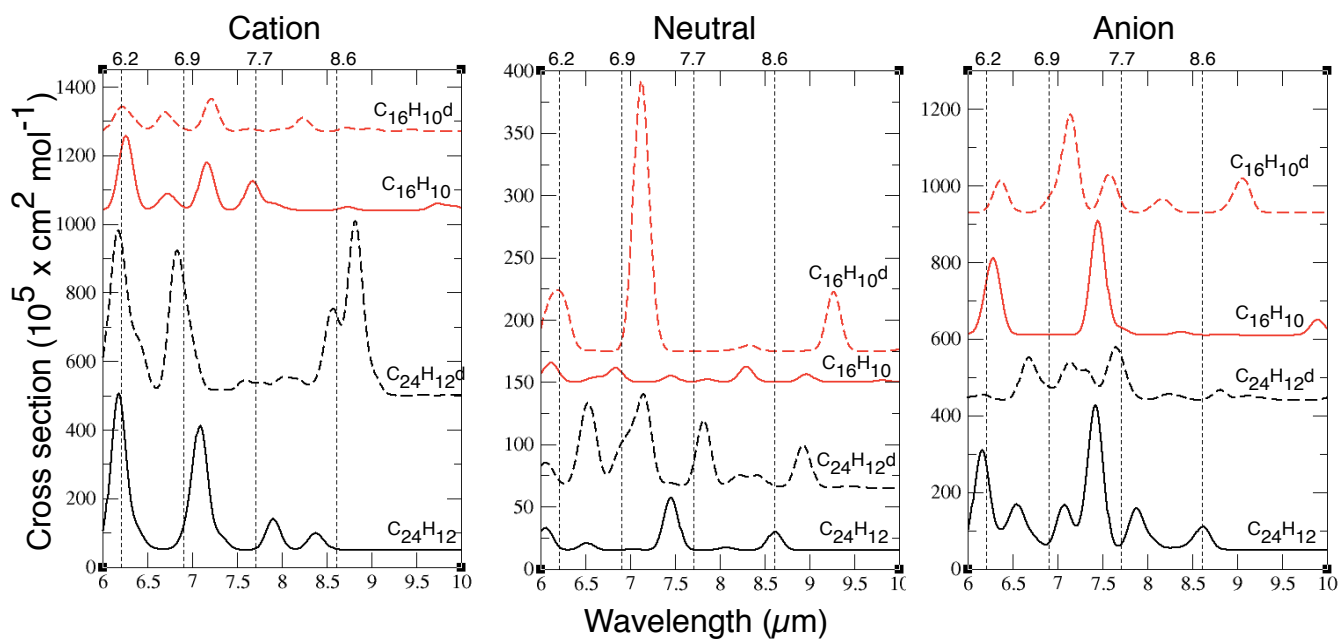}
      \caption{Theoretical IR spectra of coronene and pyrene and the defected forms in the region of 6 - 10$\mu$m. Cations, neutrals and anions are shown from left to the right. The calculations were performed using the B3LYP density functional. The changes in intensity and shifts in position of the bands are clearly visible. UIR bands at 6.2 $\mu$m, 6.9 $\mu$m, 7.7 $\mu$m and 8.6 $\mu$m are indicated by dashed lines.}
         \label{69.fig}
\end{figure*}

\begin{figure*}[!ht]
\centering
\includegraphics{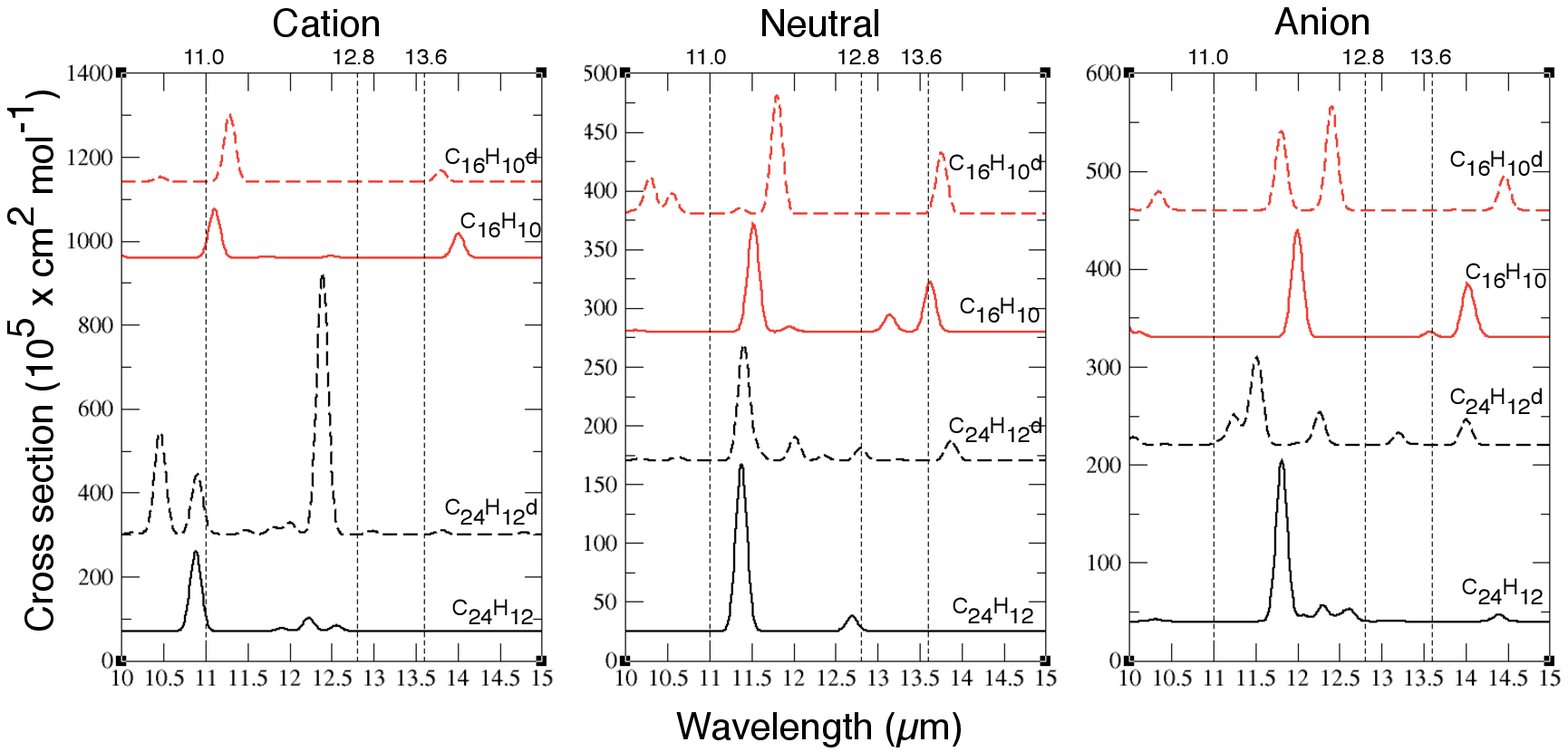}
      \caption{Theoretical IR spectra of coronene and pyrene and the defected forms in the region of 10 - 15$\mu$m. Cations, neutrals and anions are shown from left to the right. The calculations were performed using the B3LYP density functional. The changes in intensity and shifts in position of the bands are small, but still visible. UIR bands at 11.0 $\mu$m, 12.8 $\mu$m, and 13.6 $\mu$m are indicated by dashed lines.}
         \label{1015.fig}
\end{figure*}

\begin{figure*}[!ht]
\centering
\includegraphics{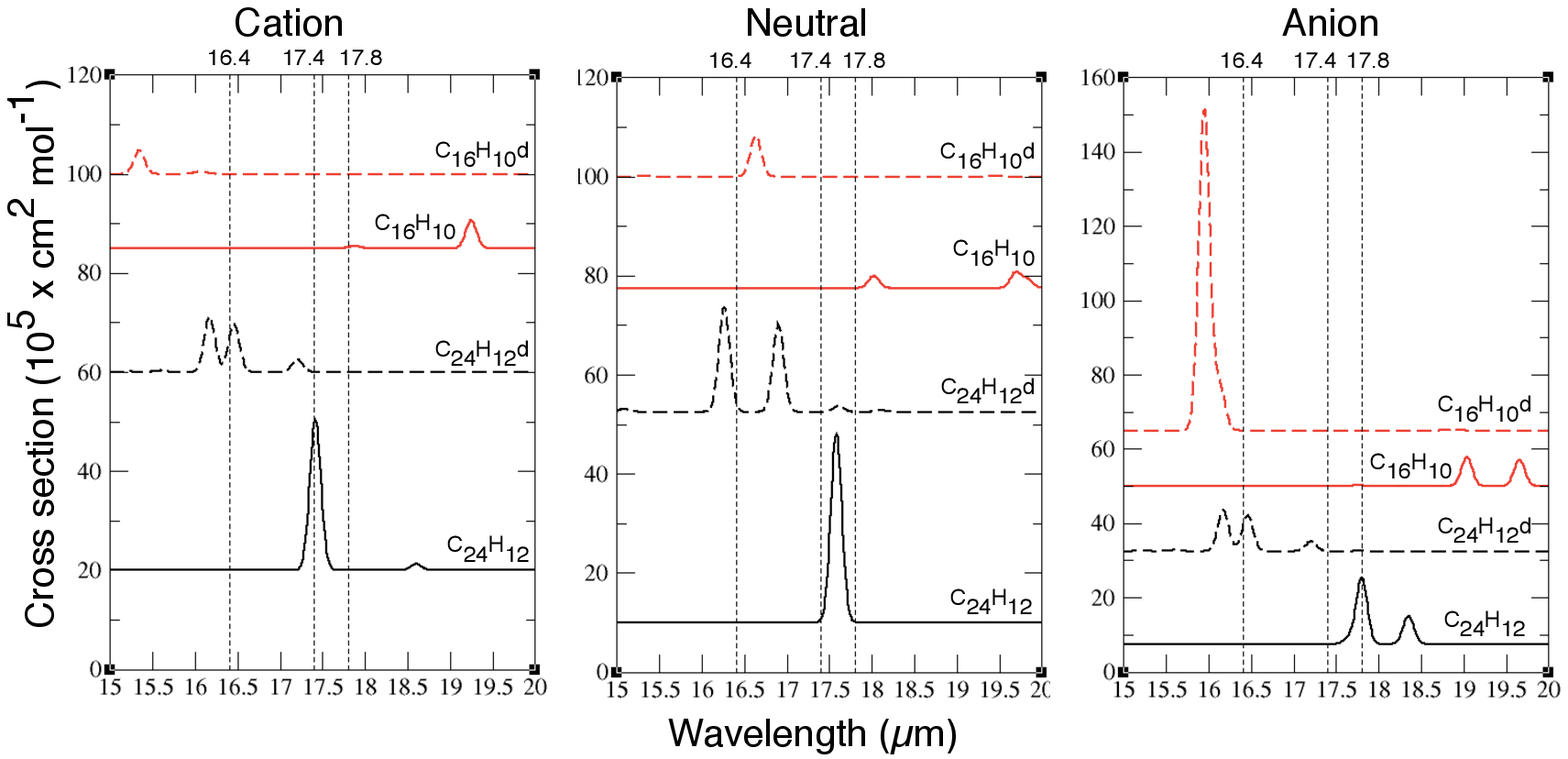}
      \caption{Theoretical IR spectra of coronene and pyrene and the defected forms in the region of 15 - 20$\mu$m. Cations, neutrals and anions are shown from left to the right. The calculations were performed using the B3LYP density functional. There are hardly any similarities between normal and defected form. UIR bands at 16.4 $\mu$m, 17.4 $\mu$m, and 17.8 $\mu$m are indicated by dashed lines.}
         \label{1520.fig}
\end{figure*}

\begin{figure}[ht!]
   \centering
   \includegraphics[width=8.8cm]{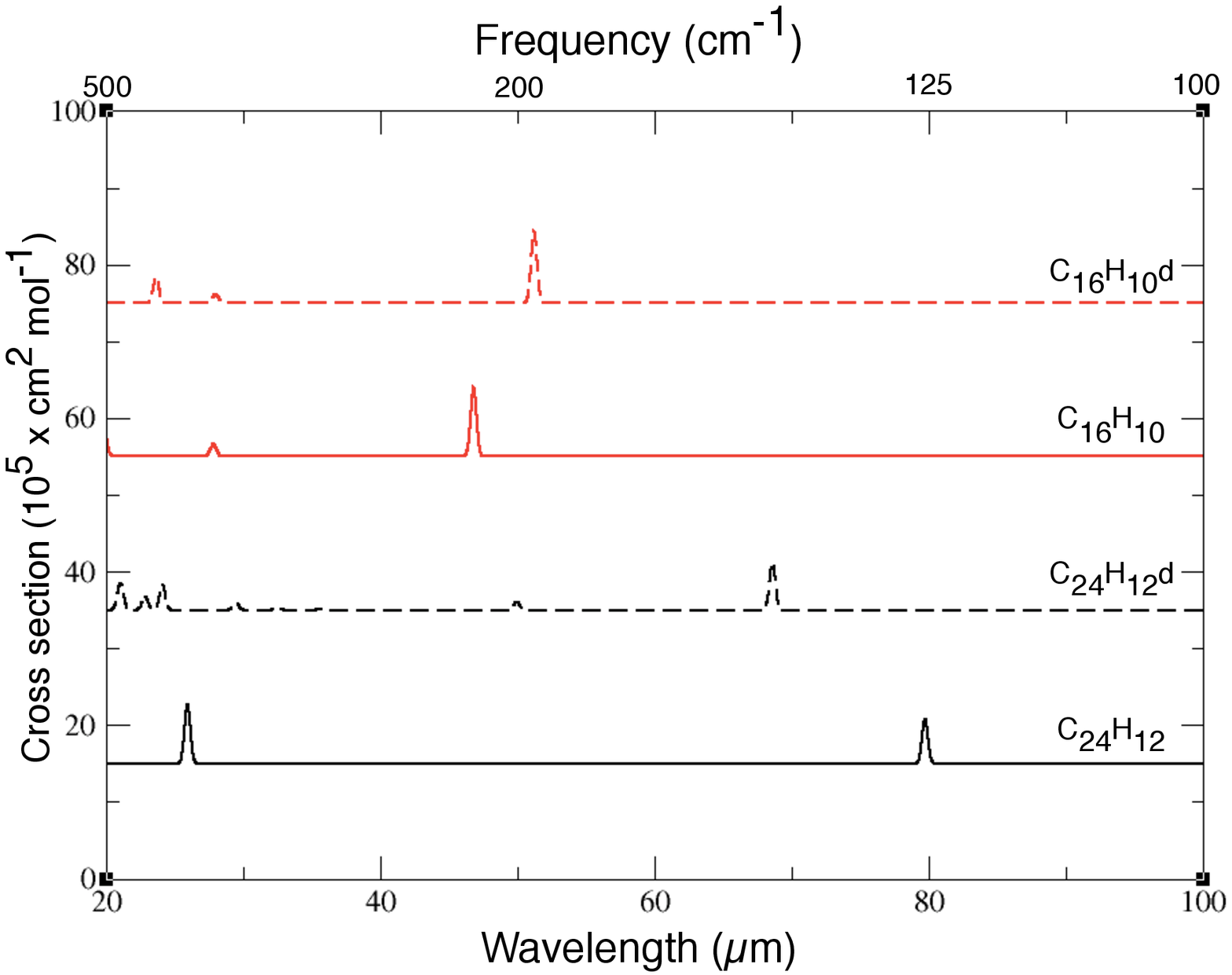}
      \caption{Theoretical IR spectra of coronene and pyrene and the defected forms in the region of 20 - 100$\mu$m. The calculations were performed using the B3LYP density functional. Additionally there is the cm$^{-1}$ axis on the top.  The changes in intensity and shifts in position of the bands are clearly visible.}
         \label{20100.fig}
\end{figure}

The resulting spectra of both coronene and pyrene as well as for the defected state of the molecules and their cations and anions are given in Figures \ref{69.fig}, \ref{1015.fig}, \ref{1520.fig} and \ref{20100.fig}. All spectra of the unmodified coronene and pyrene were compared with the NASA Ames PAH IR Spectroscopic Database (Bauschlicher et al. \cite{bausch10}). We get exactly the same IR spectra for both molecules, which is at least an indication for the reliability of all our computed data.

The first part of the spectra is the 6\ -\ 10$\mu$m region, shown in Fig. \ref{69.fig}. In this region, the vibrations are in accordance with C–C stretches and C–H in-plane bends. Especially in the 6\ -\ 10$\mu$m area it is thought that the influence of ionization is very large, the IR emission can increases sharply. The emission originates from vibrationally excited cations, proceeding via electronic excitation and intersystem crossing to highly excited vibrational levels of the ground state (Allamandola et al. \cite{alla}). Our spectra follow this trend, the cation as well as the anion show a clearly increase in intensities compared to the neutral molecule. The cations show a peak near the 6.2$\mu$m band, no matter if defected or not. But the typical bands at 7.7$\mu$m and 8.6$\mu$m cannot be found for our molecules. Only the defected form of coronene shows peaks near these bands. In general, for cations, neutrals and anions, the intensities of the defected molecules seem to exceed the results of the normal molecules. The positions of the lines are shifted in the defect state. The addition of defects leads to more bands than found in the normal molecules. For better comparison with the UIR bands, dashed lines are drawn at 6.2 $\mu$m, 6.9 $\mu$m, 7.7 $\mu$m and 8.6 $\mu$m.

In Fig. \ref{1015.fig} the 10\ -\ 15 $\mu$m region of the spectra is shown. The lines in this area originate from C-H out-of-plane bending motion. It is expected that the undefected molecules show bands near 11.2$\mu$m, wherein the peaks of the cations occur at shorter wavelengths, of the anions at longer wavelengths. Our spectra confirm this, even the molecules with defect follow this trend. We note that the differences between defected and defect-free molecules is not that big than in the 6 - 10$\mu$m region, but still not negligible. The cationic molecules still have clearly higher intensities compared to the other ones in this region. Again, for better comparison with the UIR bands, dashed lines are drawn at 11.0 $\mu$m, 12.8 $\mu$m, and 13.6 $\mu$m.

Fig. \ref{1520.fig} shows the 15\ -\ 20 $\mu$m region of the spectra. In this region, the intensities strongly decrease, but still cations and anions have higher intensities than the neutrals. The changes by adding the defects makes big differences, the molecules with defects have no similarities with the molecules without defects. There seems to be a charge effect on the 16.4$\mu$m UIR band of the defected PAHs, it can be seen for the defected coronene, whereas pyrene has a strong peak at smaller wavelength. Dashed lines are drawn at 16.4 $\mu$m, 17.4 $\mu$m, and 17.8 $\mu$m as UIR bands at these positions have been thought to originate from PAHs.

The 20\ -\ 100 $\mu$m region of the spectra in the far-IR is shown in Fig. \ref{20100.fig}. Only the neutral species is included, as the influence of ionization is very small at those wavelengths. Also in this region the changes by the defects in intensities and positions of the lines can be found.

Thus we notice in all areas of the spectra more significant deviations as in previous works. Yu \& Nyman (\cite{yu}) has raised the suspicion that there is a size dependence for the effects of the (5, 7)-member ring defect on the IR spectra of PAHs. The impact of such defects could increase with decreasing size of the PAHs. Also Ricca et al. (\cite{ricca}) found that the differences appear to become smaller as the size of the PAH increases. Boersma et al. (\cite{boersma2}) and Ricca et al. (\cite{ricca2}) discuss the difficulties of drawing any definitive conclusions about the specific PAH species that might be responsible for the UIR bands attributed to the PAHs. Our investigations represent the limiting cases with respect to the size of the PAHs. The investigation of these small systems is assumed to be an important step to a significant understanding of the role of the (5,7)-member ring defects for IR spectra of PAHs.

\subsection{Energetics}
\begin{figure}[ht!]
   \centering
   \includegraphics[width=8.8cm]{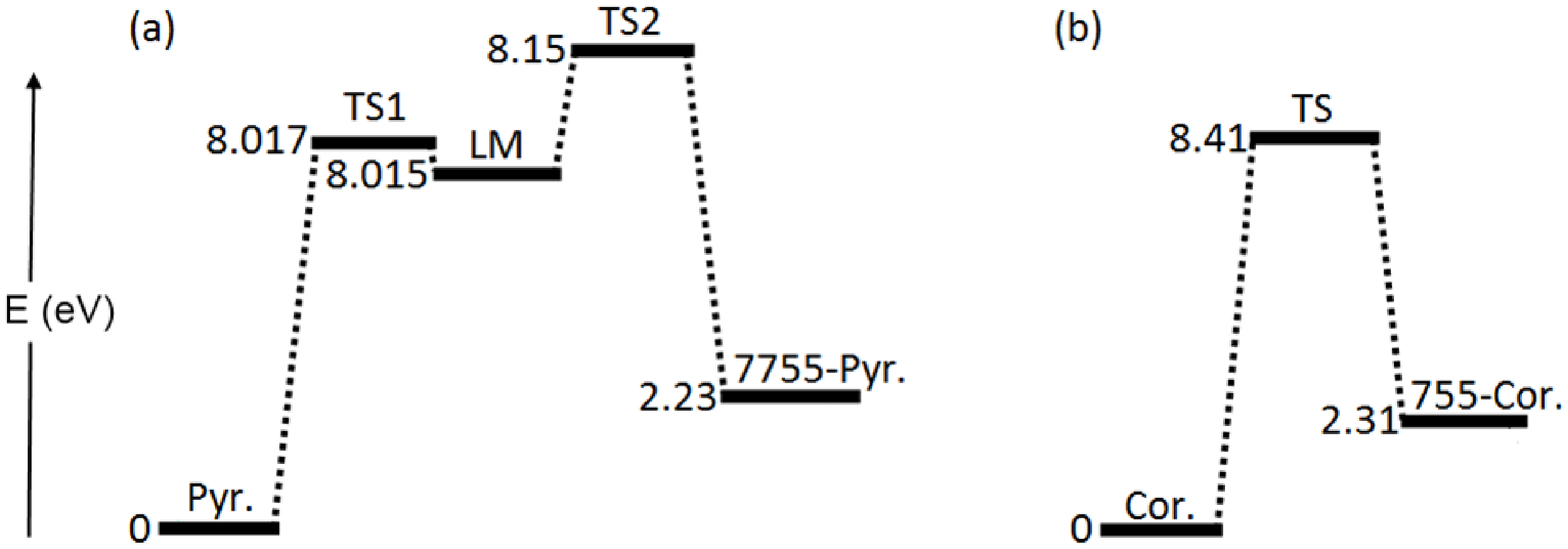}
      \caption{Relative energies (in eV) for the structures along the minimum energy pathway connecting defect-free and defected (a) pyrene and (b) coronene. Note that in case of pyrene two transition structures are involved (TS1 and TS2) separated by a shallow local minimum (LM), see also Fig. \ref{Entstehung.fig}. In contrast, the barrier separating coronene and the (5,7)-ring defected coronene corresponds to a Stone-Wales-transition structure, see also Fig. \ref{Entstehung.fig}.}
         \label{trans_scheme.fig}
\end{figure}

Table \ref{energies} summarizes the ground state energies and the energy difference between (5,7)-member ring defected and ground states. Adiabatic ionization energies and electron affinities as obtained at the B3LYP/4-31G level of theory for pyrene and coronene are shown in table \ref {ioni}. Literature values for the ionization energies and electron affinities of pyrene and coronene are 7.7, 7.21, 0.59 and 0.47 eV, respectively (Kazakov et al. \cite{kazakov}; Schroeder et al. \cite{schroeder}; Chen \& Cooks \cite{chen}; Duncan et al. \cite{duncan}). By comparison it turns out that both are underestimated by 0.5\ -\ 1 eV for both molecules. This is not surprising since the 4-31G basis set is far from being complete. For the energy difference between ground and defected states we note, however, that the estimates yielded by the B3LYP/4-31G model chemistry are close to the ones resulting from a more accurate treatment using G4(MP2) extrapolations in order to investigate the minimum energy reaction paths on the potential energy surfaces (PES) for the neutral species. The use of these accurate calculations results in energy differences of 2.23 eV in case of pyrene and 2.31 eV in case of coronene. The reaction pathways on the PES connecting ground and defected states are schematically depicted in Fig. \ref{trans_scheme.fig} for the two neutral molecules pyrene and coronene. In case of pyrene we do not observe a Stone-Wales transition leading to the Stone-Wales defect, but a more complicated path via two vastly nonplanar transition states (TS1 and TS2 in Figs. \ref{trans_scheme.fig} and \ref{Entstehung.fig}) and a shallow local minimum (LM) configuration in between (LM in Figs. \ref{trans_scheme.fig} and \ref{Entstehung.fig}) that does almost not differ from the two transition structures, see Fig. \ref{Entstehung.fig}. To find this pathway it has been necessary to replace the usual 6-31G(2df,p) basis set (Ditchfield et al. \cite{ditch}) used in G4(MP2) theory by the a bit larger cc-pVTZ basis set, because with the former one it has not been possible to achieve convergence for the local minimum structure. Furthermore, it turns out that the geometry that is achieved when the central C-C bond in pyrene is just rotated by 45 degrees corresponds to a transition structure of second order exhibiting two imaginary frequencies. In case of coronene we observe that the transition between the ground and defected state actually proceeds via the Stone-Wales transition, however, obvious from the structure of the molecule, not resulting in a Stone-Wales defect (see Fig. \ref{Entstehung.fig}). The minimum activation barriers for transforming pyrene and coronene into their (5,7)-ring defected forms are found to be 8.21 and 8.41 eV, respectively. The corresponding reverse barriers are 5.92 and 6.10 eV, respectively, and are thus close to the 6.47 and 6.41 eV found for the neutral and cationic $C_{48}H_{18}$ PAH model studied in an earlier investigation (Yu \& Nyman \cite{yu}). Hence, the energy required to build (5,7)-ring defects seems to be insensitive to the actual size of the PAH molecule which is in line with the explanation given for the high barriers in the previous study of Yu \& Nyman (\cite{yu}). Therein, it was shown that for the transformation two $\pi$ C-C bonds have to be broken which requires a large amount of energy. The latter should obviously be quite insensitive to the size of the molecule.

\begin{table}
\caption{Ground state energies and energy differences for neutral (q=0), anionic (q=-1) and cationic (q=+1) species.} 
\label{energies} 
\centering 
\begin{tabular}{c c c c} 
\hline\hline 
species &  & energy [au] & \\ 
 & $q = -1$ & $q = 0$ & $q = +1$ \\
\hline 
coronene & $-920.70323$ & $-920.702546$ & $-920.452811$ \\
755-coronene & $-920.64816$ & $-920.604581$ & $-920.360825$ \\
pyrene & $-614.974253$ & $-614.977292$ & $-614.725558$ \\
7755-pyrene & $-614.90863$ & $-614.892064$ & $-614.651359$ \\
\hline 
\hline
 & & $\bigtriangleup$ E [eV] & \\
 & $q = -1$ & $q = 0$ & $q = +1$ \\
\hline
coronene & & & \\
755-coronene & $1.50$ & $2.67$ & $2.50$ \\
pyrene & & & \\
7755-pyrene & $1.79$ & $2.32$ & $2.02$ \\
\hline
\end{tabular}
\tablefoot{The ground state energies and the energy differences for coronene and pyrene and their (5,7)-ring defected forms are given as obtained at the B3LYP/4-31G level of theory.}
\end{table}

\begin{table}
\caption{Adiabatic ionization energies and electron affinities.} 
\label{ioni} 
\centering 
\begin{tabular}{c c c} 
\hline\hline 
species & ionization energy [eV] & electron affinity [eV] \\
 & $q = 0$ & $q = 0$ \\
\hline
coronene & $6.80$ & $0.02$ \\
755-coronene & $6.63$ & $1.19$ \\
pyrene & $6.85$ & $(0.14)$\\
7755-pyrene & $6.55$ & $0.45$ \\
\hline 
\end{tabular}
\tablefoot{The adiabatic ionization energies and electron affinities for coronene and pyrene and their (5,7)-ring defected forms corresponding to the B3LYP/4-31G model chemistry. The value for the electron affinity of pyrene is taken from the calculations using BP86, since B3LYP yields no energetically stable anion (see Table \ref{energies}).}
\end{table}

\section{Discussion}
\subsection{Formation}

\begin{figure*}[!ht]
\centering
\includegraphics{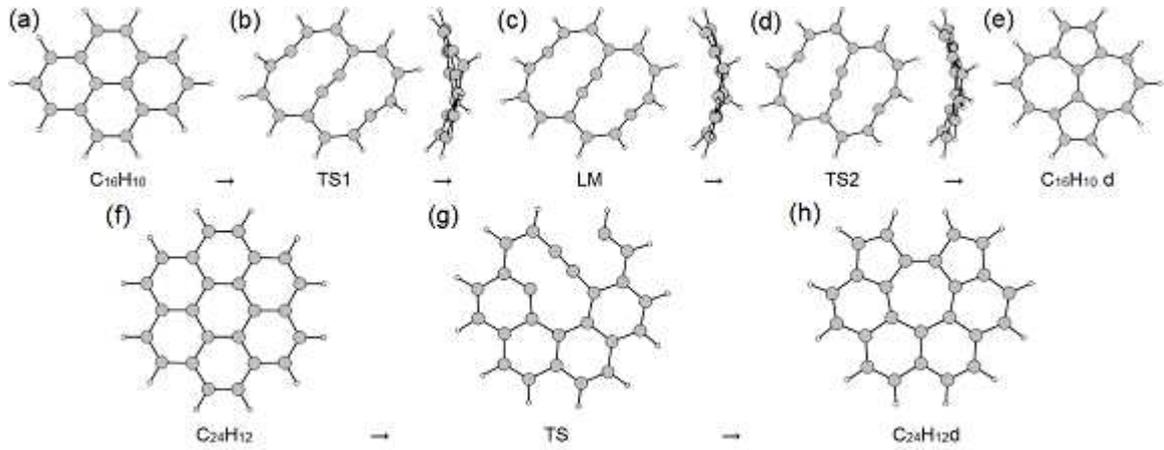}
\caption{Possible formation pathways of the (5,7)-member ring defects depicted for (a)-(e) pyrene and (f)-(h) coronene.}
\label{Entstehung.fig}
\end{figure*}

The environment of carbon rich AGB stars seems to be the perfect place for PAH formation (Allamandola et al. \cite{alla}). One of the most important molecules in carbon-rich environments is $C_{2}H_{2}$, which is presumably the parent molecule in PAH formation (Allamandola et al. \cite{alla}). A detailed study on the growth of carbon molecules was done by Whitesides \& Frenklach (\cite{white}), where they also discussed the role of 5-member rings.

The defect may not only be formed in the growing process of the PAHs. Also transitions between the ground state and the defected form might play a role (see Fig. \ref{Entstehung.fig}). As we have discussed in section 3.2, the minimum activation barriers for transforming pyrene and coronene into their (5,7)-member ring defected forms is 8.21 eV for pyrene and 8.41 eV for coronene, respectively. The energy required to build (5,7)-member ring defects seems to be insensitive to the size of the PAH molecule. We will now discuss how these barriers can be overcome. A very recent study of Champeaux et al. (\cite{champ}) discussed the physical interactions of PAHs with stellar particle radiation, especially the interaction of coronene with protons. An interaction will not necessarily destroy or fragment the molecules. If the particle has enough energy and hits exactly on a knot of the molecule, a change of the ring structure from 6-membered rings into pentagons and heptagons may occur (Huber et al. \cite{huber}). The particle has to have at least enough energy to overcome the activation barrier. Clearly not every hit with sufficient energy will cause a deformation, but also maybe ionization, dissociation or hydrogenation. Too much energy will lead to the destruction of the molecule. The spatial angle for hit should be around $90^{\circ}, \pm 10^{\circ}$, but this parameter was not investigated in detail, and may also vary. Possible candidate particles can be found in different environments like stellar winds, AGB stars and winds or PNe. The probability for a collision in such environments can be estimated as follows: Using energy conservation and momentum conservation and taking the parameters of a typical C-rich AGB stellar wind, the collision probability can be calculated. Even these very simplified calculations show a few hits per second for a simple stellar wind up to a distance of about 20 R$_\odot$. Recent observations showed evidence of mixed chemistry with emission from both silicate dust and PAHs (Guzman-Ramirez et al. \cite{liz1}) or the simultaneous presence of mid-IR features attributed to neutral fullerene molecules (i.e. $C_{60}$) and PAHs (Garc\'{\i}a-Hern\'{a}ndez et al. \cite{garcia}). A very recent study discussed that PAHs are located and even formed at the outer edge of dense tori (Guzman-Ramirez et al. \cite{liz2}). The presence of a dense torus has been strongly associated with the rich and mixed chemistry seen in these regions. All these observed environments strongly enable the transition from the ground state to the defect state of the molecules.

\subsection{Observations}
\begin{figure}[ht!]
   \centering
   \includegraphics[width=8.8cm]{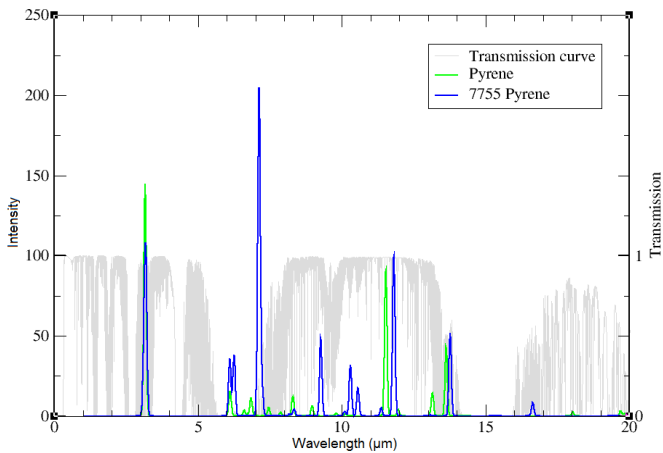}
      \caption{A comparison between the IR spectra of pyrene (green) and pyrene exhibiting the (5,7)-member ring defect (blue), both neutral. The changes in intensity and shifts in positions of the lines in the spectrum are clearly visible. In grey we show the transmission curve of the atmosphere, to conditions as are customary for Paranal.}
         \label{mex1.fig}
\end{figure}

\begin{figure}[ht!]
   \centering
   \includegraphics[width=8.8cm]{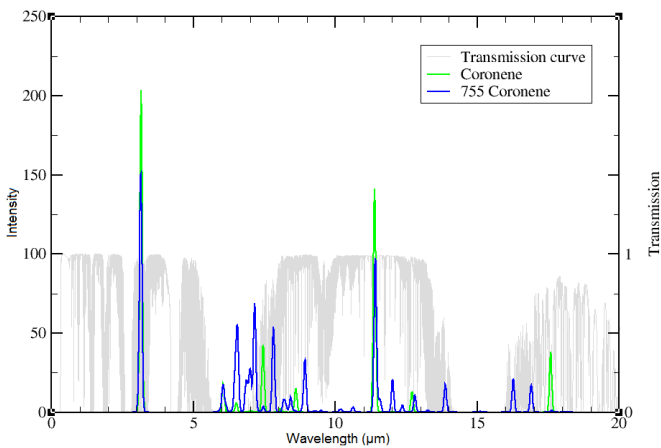}
      \caption{A comparison between the IR spectra of coronene (green) and coronene exhibiting the (5,7)-member ring defect (blue), both neutral. Same as for pyrene, the lines change in intensity and position within the spectrum. In grey again the transmission curve of our atmosphere, to get an idea about the observability and detectability.}
         \label{mex2.fig}
\end{figure}

To get an idea about the possibilities to observe the effects of the (5,7)-member ring defects, Fig. \ref{mex1.fig} and \ref{mex2.fig} show a comparison of the defected molecule with the normal one in the region from 0\ -\ 20 $\mu$m. In addition the transmission curve of the Earth's atmosphere is plotted, calculated with the code LNFL/LBLRTM (Clough et al. \cite{clough}) for airmass = 1 and a years averaged atmospheric profile scaled to 3mm PWV (precipitable water vapor), a typical PWV value for Paranal. So one can get an impression of best detectable lines and bands best suited for a study. Fig. \ref{mex1.fig} shows pyrene and its defected form, Fig. \ref{mex2.fig} shows coronene and its defected form, all of them in their neutral states.

\section{Conclusions}
PAHs observed in astrophysical environments are thought to be formed mainly in the surroundings of C-rich AGB stars. The growth of the aromatic hydrocarbon rings includes effects of 5-member and 7-member ring defects (Whitesides \& Frenklach \cite{white}). The still unidentified IR features seen in the observations require further computational, laboratory and observational investigations. We have performed a theoretical study of the IR spectra of PAHs containing (5, 7)-member ring defects. Using density functional theory, we investigated the effect of such defects on the IR spectra of pyrene ($C_{16}H_{10}$) and coronene ($C_{24}H_{12}$) and their cations and anions. In addition, we explore parts of the potential energy surface of the neutral species.

The addition of (5, 7)-member ring defects into the skeleton of these small PAH molecules has an astonishing large effect on the IR spectra. Pyrene and coronene exhibiting a (5, 7)-member ring defect lose their typical spectroscopic signature. There is a clear difference in the spectrum of the unmodified molecule and the one exhibiting the (5, 7)–member ring defect. Coronene and pyrene are amongst the smallest PAHs which may exhibit such ring defects. This implies that the internal heat quantity taken by other parts after a hit is the smallest in these molecules, and therefore the probability of reaching the energy of the barrier is high. In this study we represented the boundaries in terms of the size of the PAHs. The minimal energy pathways between the ground and (5, 7)–member ring defected states exhibit minimum activation barriers of 8.21 eV for pyrene and 8.41 eV for coronene for the transformation. The corresponding reverse barriers are 5.92 eV for pyrene and 6.10 eV for coronene, which agrees well with the results of Yu et al. \cite{yu}. The size of the PAH molecule seems to have no influence on the energy required to build (5, 7)-ring defects. Whereas pyrene retains some of its symmetry due to the symmetry exhibited by the Stone-Wales defect itself, coronene loses much more of its symmetry. The formation of these (5, 7)-ring defects in PAHs may be well supported in AGB stars or PNe. Those environments strongly enable the transition from the ground state to the defect state. Therefore the knowledge of the IR spectra of these molecules can facilitate future investigations in understanding the unidentified IR emission bands.
To study the (5, 7)-membered ring defects in other PAH molecules is believed to be an important step to a significant understanding of the UIR bands and the IR spectra of PAHs, although the defects in astronomical PAH molecules will be extremely difficult to detect with current instruments and tools. But future telescopes and instruments are promising a chance towards the observability of these molecules and their defects. High resolution IR spectrographs of the European Extremely Large Telescope (E-ELT) type telescopes will allow a spatial and wavelength segregation from telluric lines in even highly obscured wavelengths domains (e.g., METIS-IFU with a spectral resolution of R $\sim$ 100 000, Brandl et al. (\cite{brandl})). The James Webb Space Telescope (JWST) will allow to complete studies of the bright stars observed the IR with extremely high S/N ratios(e.g., MIRI at the JWST, Wright et al.\cite{wright}).

\begin{acknowledgements}
      S.{\"O}. and S.E.H. acknowledge support by the Austrian \emph{Fonds zur Wissenschaftlichen Forschung, FWF\/} DK$+$ project Computational Intersiciplinary Modelling W1227-N16. This work was supported by the Austrian Ministry of Science BMWF as part of the UniInfrastrukturprogramm of the Focal Point Scientific Computing at the University of Innsbruck.
\end{acknowledgements}

\end{document}